\documentclass[]{aastex62} 
\usepackage{amsmath,amstext}

\newcommand{\be}{\begin{eqnarray}}
\newcommand{\ee}{\end{eqnarray}}
\usepackage{xcolor}
\usepackage{comment}

\newcommand{\MSun} {\mbox{$M_{\odot}$}}

\received{\today}
\submitjournal{ApJ}

\shorttitle{Close stellar flybys common in low-mass clusters}
\shortauthors{Pfalzner \& Govind}

\begin{document}

\title{Close stellar flybys common in low-mass clusters}

\author[0000-0002-5003-4714]{Susanne Pfalzner} 
\affiliation{J\"ulich Supercomputing Center, Forschungszentrum J\"ulich, 52428 J\"ulich, Germany}
\affiliation{Max-Planck-Institut f\"ur Radioastronomie, Auf dem H\"ugel 69, 53121 Bonn, Germany}

\author[0000-0002-6652-0303]{Amith Govind}
\affiliation{J\"ulich Supercomputing Center, Forschungszentrum J\"ulich, 52428 J\"ulich, Germany}
\affiliation{Max-Planck-Institut f\"ur Radioastronomie, Auf dem H\"ugel 69, 53121 Bonn, Germany}

\email{s.pfalzner@fz-juelich}

\begin{abstract}
Numerous protoplanetary discs show distinct spiral arms features. While possibly caused by a range of processes, detailed pattern analysis points at close stellar flybys as cause for some of them. Surprisingly, these discs reside in young low-mass clusters, where close stellar flybys are expected to be rare. This fact motivated us to take a fresh look at the frequency of close flybys in low-mass clusters. In the solar neighbourhood, low-mass clusters have smaller half-mass radii than their more massive counterparts. We show that this observational fact results in the mean and central stellar density of low-mass clusters being approximately the same as in high-mass clusters, which is rarely reflected in theoretical studies. We perform N-body simulations of the stellar dynamics in young clusters obeying the observed mass-radius relation.  Taking the mean disc truncation radius as a proxy for the degree of influence of the environment, we find that the influence of the environment on discs is more or less the same in low- and high-mass clusters. Even the fraction of small discs ($<$ 10 au) is nearly identical. Our main conclusion is that the frequency of close flybys seems to have been severely underestimated for low-mass clusters. A testable prediction of this hypothesis is that low-mass clusters should contain  10\%--15\% of discs smaller than 30 au truncated by flybys. These truncated discs should be distinguishable from primordially small discs by their steep outer edge.
\end{abstract}


\section{Introduction} 
\label{sec:intro}

The observed diversity in the structure of protoplanetary discs \citep[for example, ][]{Andrews_2020,vandermarel_2021} and the planetary systems \citep[for example,][and references therein]{Kaltenegger:2017,Miguel:2020}, triggers a renewed interest in the question of the effect of the environment on planet formation \citep[recent examples being, ][]{Cai:2018,Winter:2020,Adibekyan_2021,Dai:2021}. One way how the environment influences a forming planetary system is close stellar flybys. These stellar flybys can potentially truncate or even completely destroy the protoplanetary discs, thus influencing the total size and frequency of planetary systems \citep[for example, ][]{Scally_n_Clarke_2001,Adams_2006,Olczak_2006,Steinhausen:2014,Rosotti_2014,Portegies_Zwart_2016,Vincke_2016,Concha:2019,Jimenez:2020,Concha:2021}. Close stellar flybys are expected to be most common in the central areas of high-mass clusters but rare in low-mass associations. Finding signs of the stellar flybys seemed for a long time elusive, as only a handful of objects showed the characteristic spiral arm patterns. The situation changed when the ALMA telescope allowed for the first time to produce highly resolved imaging of protoplanetary discs. Suddenly, characteristic spiral structures are abundantly observed  \citep{Meru:2017,Long:2018,Dong:2018,Andrews:2018}. Spiral structures can either be induced by a companion or a stellar flyby, so that undetected companions can be the reason behind the pattern. However, the repeated interaction with a bound companion generates more compact spirals.

Close stellar flybys are expected to be most common in locations of high stellar density. However, the discs with spiral arm patterns are often located in low-mass young associations.   In several cases, the spiral pattern details clearly point to a flyby rather than a companion. So ongoing (or past) flyby have been suspected for RW Aur \citep{Cabrit:2006,Dai:2015,Rodriguez:2018}, AS 205 \citep{Kurtovic:2018}, HV Tau and Do Tau \citep{Winter:2018},  FU Ori \citep{Beck:2012, Takami:2018, Perez:2020}, V2775 Ori \citep{Zurlo:2017}, V1647 Ori \citep{Principe:2018}, Z CMa \citep{Takami:2018}, SU Aur \citep{Akiyama:2019} and UX Tau \citep{Menard:2020}. These systems often show other signs typical for a close stellar flyby, such as disc truncation and warping accompanied by dimming and outburst events \citep{Cuello:2019}. Assuming that close stellar flybys indeed induced the spiral pattern, it would mean that close flybys in low-mass associations are more common than generally believed. Here, we test this hypothesis by performing cluster simulations based on the observed properties of young ($<$ 3 Myr) stellar groups typical for the solar neighbourhood.

First, we motivate why close encounters might be more common than expected in low-mass clusters (section \ref{sec:low_mass_cluster_properties}). The central argument here is that low-mass clusters have a smaller size than high-mass clusters leading to a higher mean stellar density. In Section \ref{sec:method} we describe the method we use to simulate the flyby dynamics in clusters and justify the approximations made in our toy model. The main result of our simulations is that close flybys are actually so common in low-mass clusters, that their effect on discs might be comparable to that in high-mass stellar groups. We show that our results agree with observational results of the frequency of small discs in low-mass clusters. 
Finally, we make two predictions based on our simulation results that dedicated observations could test. These predictions link to the interpretation of disc and planetary system diversity. \\

\begin{figure}[t]
\centering\includegraphics[width=0.90\textwidth]{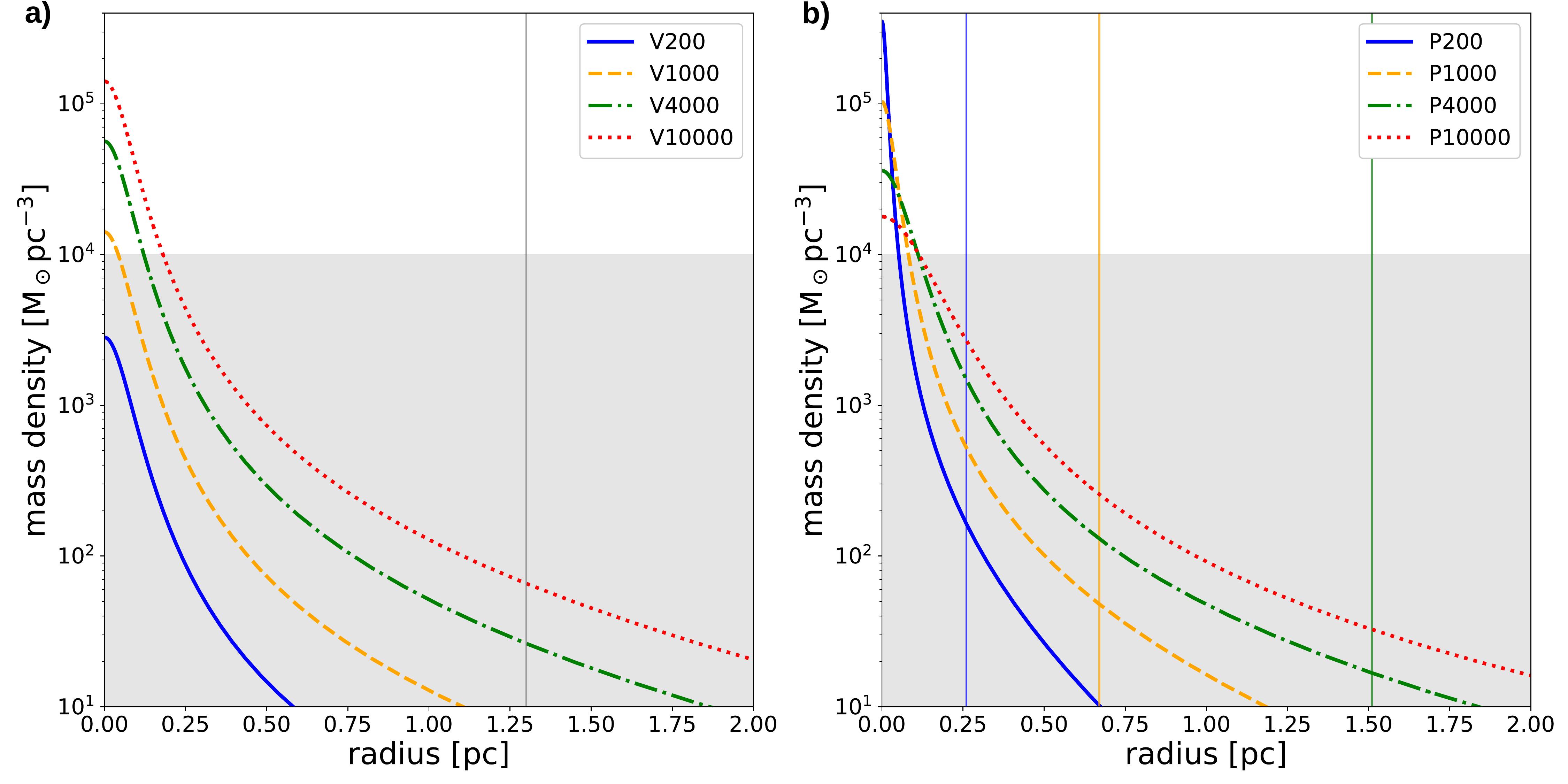}
\caption{Cluster density profiles for stellar groups consisting of N=200, 1000, 4000 and 10 000 stars distributed according to a King W=9 profile. In both figures the grey area indicates the density range where close stellar flybys are expected to be uncommon. {\bf a)} for \mbox{$R_\mathrm{c}$= 1.3 pc } correspondind to models V 200 --- V 10 000 and {\bf b)} for \mbox{$R_\mathrm{c}=(M_\mathrm{c}/C)^{1/\gamma}$} corresponding to models P 200 --- P 10 000. The vertical lines indicate the half-mass radii in the respective models. For model P 10 000 the half-mass radius is outside the shown spatial range.
}
\label{fig:King}
\end{figure}

\section{Low- vs. high-mass stellar groups}
\label{sec:low_mass_cluster_properties}

In the context of observations, it is often stated that a sizeable effect by the environment on the discs happens only in areas with local stellar number density exceeding $n$ =10$^4$/pc$^{-3}$ \citep{Guthermuth:2005}. Only here, close stellar flybys are expected to happen at a sufficiently high frequency. It is a standard argument that such high stellar densities are exclusively found in the dense central areas of high-mass clusters\footnote{In the following we use the term high-mass and high-N clusters as synonym.} containing several thousand stars \citep{Adams:2010,Malmberg:2011,Craig_2013,Dukes:2012,Winter:2020,Parker_2020}. The central area of the Orion nebula cluster (ONC) is often referred to as the prime example of such a close-by dense environment. Fig. \ref{fig:King}a) seems to confirm the expectation that only clusters with a high number of stars, and therefore mass, exceed the above noted stellar density limit.  It shows the stellar density distribution for stellar groups with different numbers of stars assuming a King profile with a half-mass radius of $R_{hm}$=1.3 pc. Only the stellar groups with $\geq$ 4000 stars reach central densities exceeding the threshold of 10$^4$/pc$^{-3}$. Based on this expectation, so far, the effect of flybys was preferentially simulated for relatively high-mass clusters containing
 $N \geq$ 1000 stars, where $N$ denotes the number of cluster members \citep[for example,][]{Scally_n_Clarke_2001,Olczak_2010,Vincke_2015,Vincke_2016,Parker_2020,Concha:2021}.

\begin{figure}[t]
\centering
\includegraphics[width=0.48\textwidth]{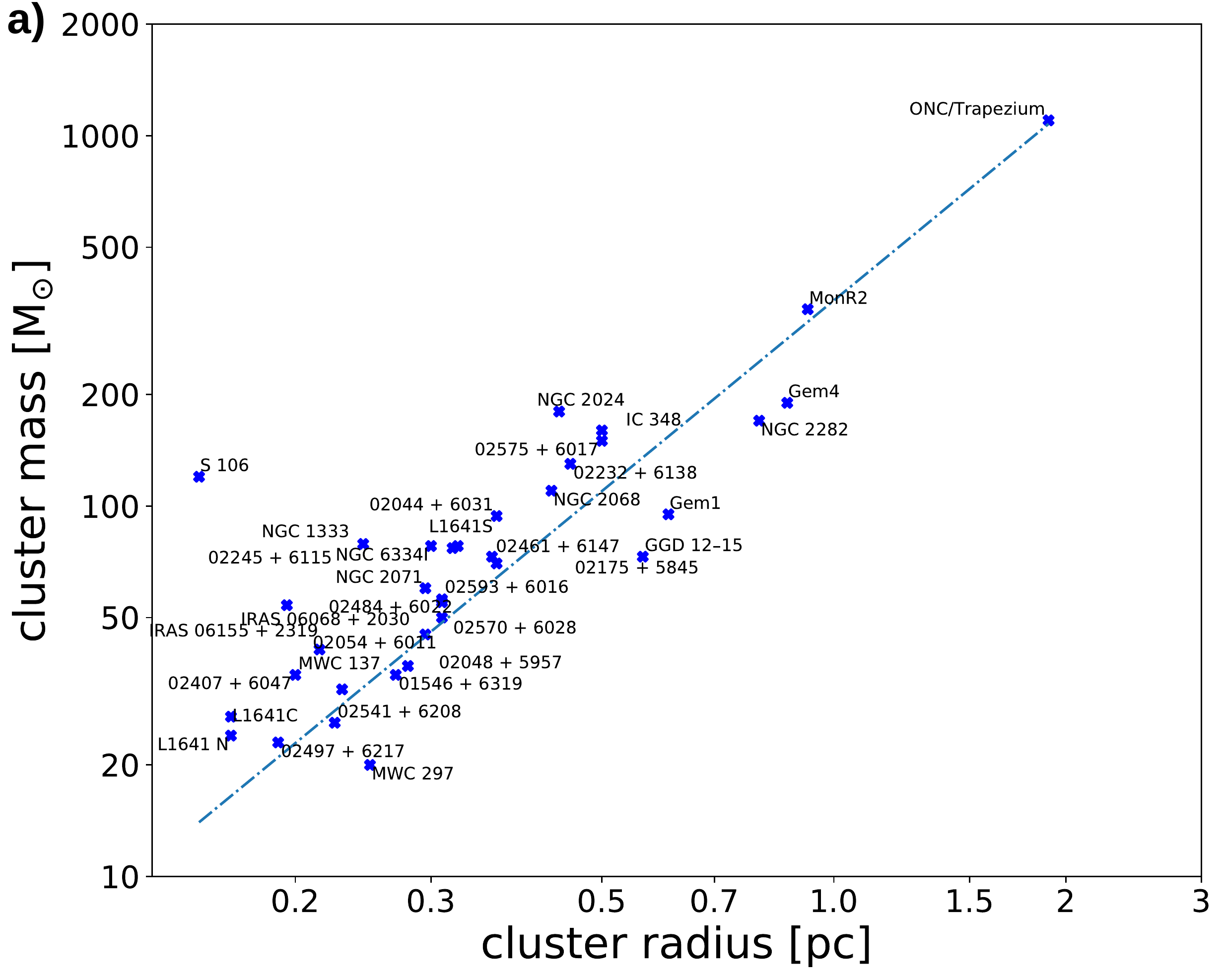}

\includegraphics[width=0.48\textwidth]{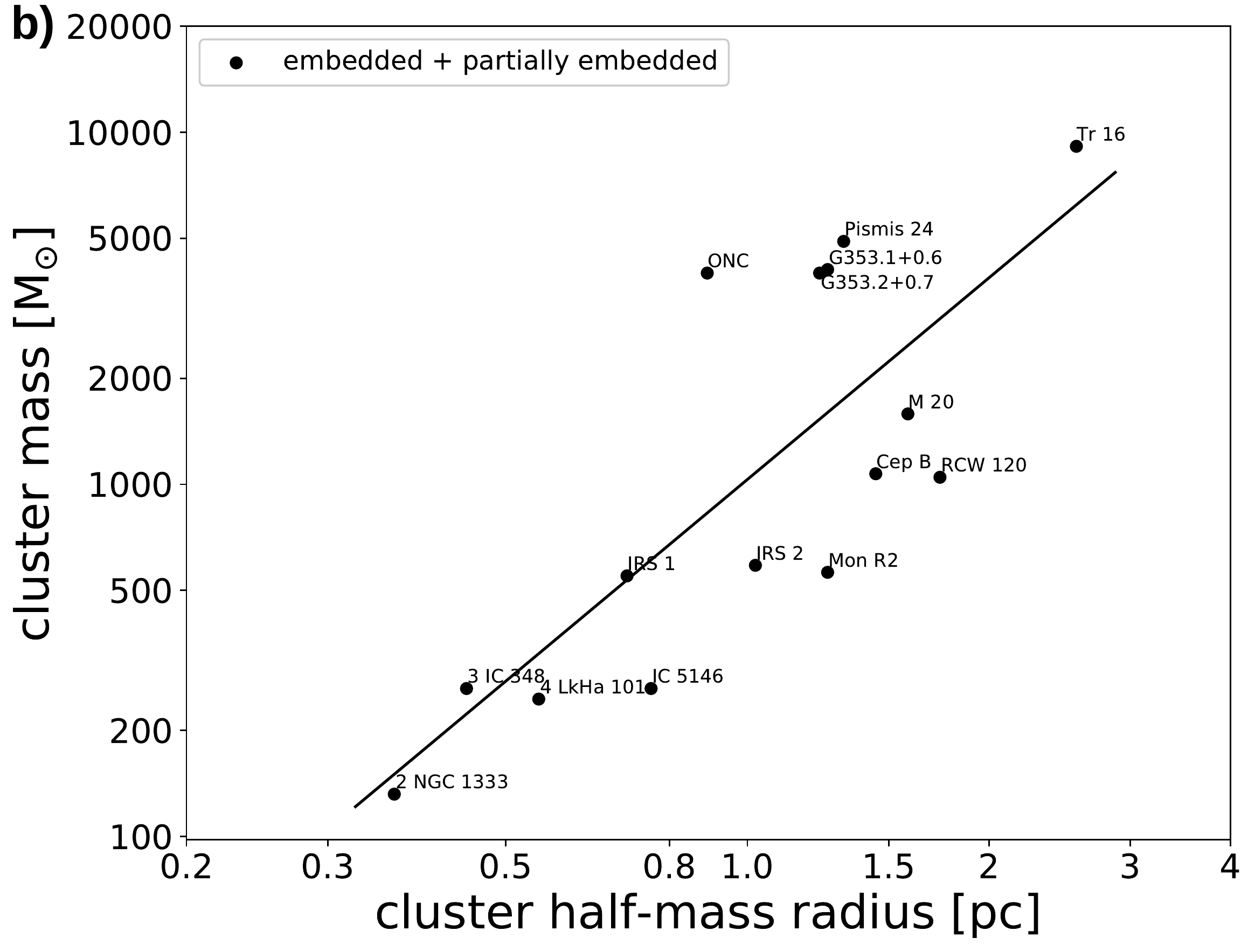}

\includegraphics[width=0.48\textwidth]{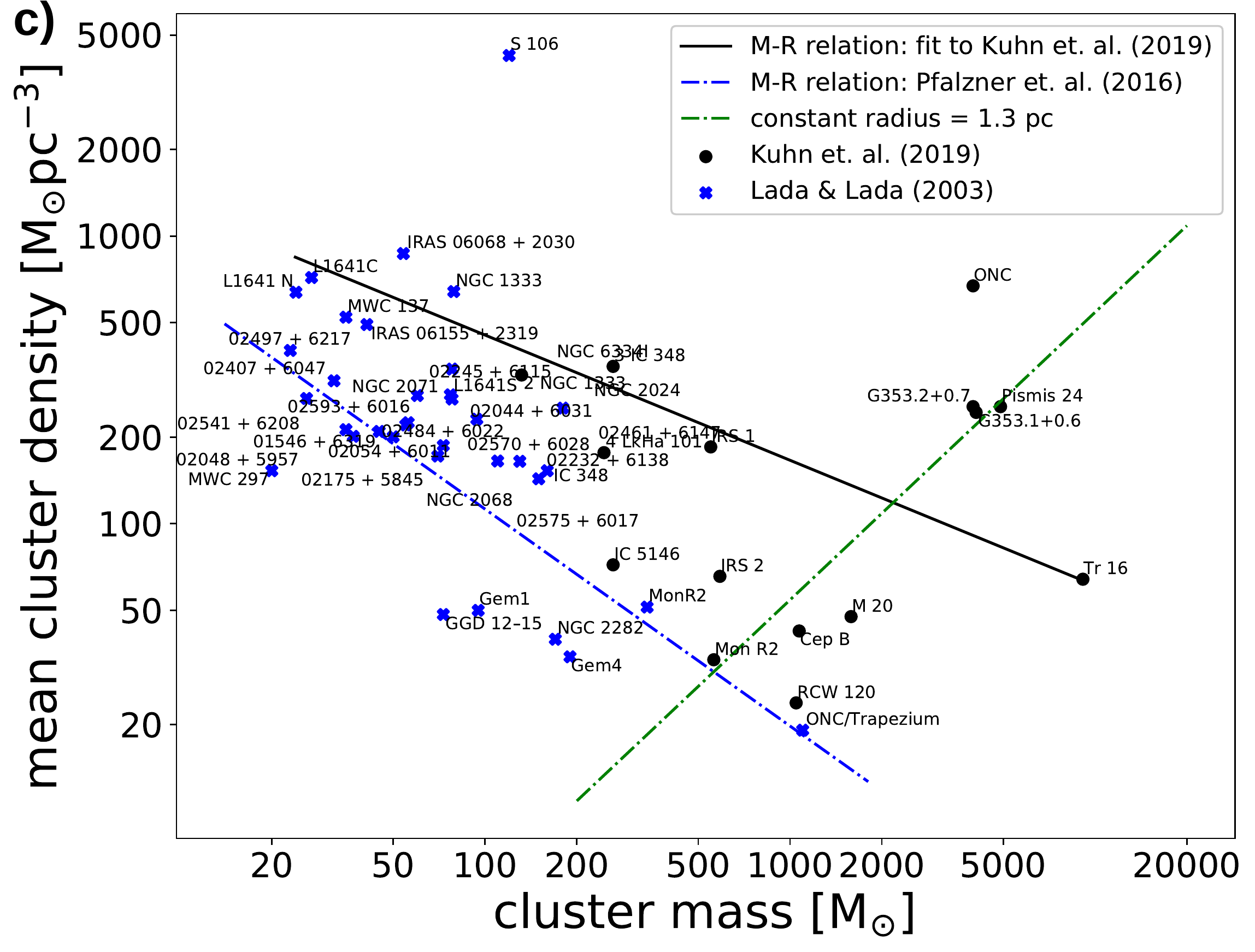}
\caption{The two top panels show the mass vs half-mass radius for stellar groups in the solar neighbourhood. a) shows a modified version of Fig.1a) in \cite{Pfalzner_2016} for the data given by \cite{lada_lada_2003}.  b) illustrates the same but for the data given in \cite{Kuhn_2019},  excluding the data for Tr14 and M17, as they will likely develop into long-lived open clusters. The bottom panel shows the corresponding mean stellar density vs estimated total cluster mass corresponding to a) as a blue dashed line and to b) as a black drawn line. The green dashed line shows the mean stellar density assuming a half-mass radius of 1.3 pc independent of the cluster mass as, for example, used in the simulations by \cite{Vincke_2016}.}
\label{fig:mass-radius-relation}
\end{figure}

However, there is a potential problem with the above argument. Only if one considers low- and high-mass stellar groups of the same size, a high-mass group always has a higher central density than a low-mass group. However, if one looks at the properties of young stellar groups in the solar neighbourhood, the equal-size assumption is not applicable. The nearby young stellar groups ($<$ 2 kpc, age $<$ 10 Myr) show a general trend that low-mass stellar groups have a smaller size than high-mass ones. This is illustrated in Fig. \ref{fig:mass-radius-relation}a) for the data given in \cite{lada_lada_2003} and in Fig. \ref{fig:mass-radius-relation}b) for the recent cluster sample from \cite{Kuhn_2019}. More examples can be found in \cite{Pfalzner_2016}.  The relation between a cluster's mass and its half-mass radius can be approximated by
\be 
M_\mathrm{c} = C R_\mathrm{c}^\gamma,
\label{eq:mass-radius}
\ee
where $M_\mathrm{c}$ denotes the cluster mass, $C$ is a constant, $R_\mathrm{c}$ the cluster half-mass radius and $\gamma$ the scaling exponent. The constant $C$ and the scaling exponent $\gamma$ typically differ somewhat between considered observational data due to observational uncertainties in the cluster masses and radii. Note, in particular, the cluster size values for the same cluster can differ considerably between different samples. The reason is methodological differences, for example,  in cluster member identification, size definitions and data being taken at different wavelengths (for a more detailed discussion on the reasons for individual differences see \citet{Pfalzner_2016}). it is essential that here standardized methods are developed soon.  However, for all investigated data compilations, the scaling exponent lies typically in the range 1.6---2.1 \citep{Pfalzner_2016}. Using the full data set by \citep{Kuhn_2019}, we obtain a best fit for $C$=1263 and $\gamma$=2.09. Relation \ref{eq:mass-radius} holds only for the relatively short-lived stellar groups in the solar neighbourhood. It does not apply to the more compact young clusters that develop into long-lived clusters located at larger distances in the Galaxy's spiral arms (see Pfalzner \& Kacmarek 2013. The data set by \citet{Kuhn_2019} contains two clusters that likely develop into long-lived clusters -- the distinctly compacter clusters Tr 14 and M 17, which also are of much higher mass than the rest of the sample. If we exclude these two clusters, the best-fit parameters change to $C$=1032 and $\gamma$ = 1.9. However, whichever {\emph homogeneous}  sample of young stellar groups in the solar neighbourhood one uses,  the general trend of an increase in half-mass radius with cluster mass always persists. 

Thus far, this trend of smaller sizes for low-mass clusters has been  largely neglected in simulations that concentrate on the effect of stellar flybys. An exception are the works by \citet{Adams:2006} who derived an  equivalent relation but expressed in terms of the number of stars in the cluster, $N$. They also applied their $N/R_\mathrm{c}$-relation to their cluster simulations \citep{Adams:2010,Adams:2014}, but mainly concentrated on the effects of external photo-evaporation and supernova enrichment, while \citet{Proszkow:2009} looked at the flyby history included low-mass clusters, but did not consider the consequences of close flybys in such environments in detail. Supernovae explosions and external photoevaporation --- require the presence of massive stars to be efficient, therefore low-mass clusters are basically unaffected by these two environmental processes. We will see in the following that the situation is completely different for the process of close stellar flybys. 

Apart from these works, it was mostly assumed that cluster mass and size are uncorrelated. Therefore, either a parameter grid of sizes and masses was investigated without connecting to the mass-size relation in young clusters or it was assumed that all clusters have approximately the same size when fully formed.  The latter approach seemed reasonable as theoretical models which backwards-engineered the initial sizes of clusters \citep{marks_kroupa_2012} indicated that the considered clusters developed all from an initial size of $\sim$ 0.1 pc independent of their mass. However, the considered clusters spanned over an age range of several hundred Myr and therefore contained a large fraction of long-lived open clusters.  As Figs.  \ref{fig:mass-radius-relation}a) and b) demonstrate,  this conclusion is not necessarily transferable to the short-lived ($<$ 10 Myr) stellar groups in the solar neighbourhood. 

Applying the relation  $\rho = 3M_\mathrm{c}/4\pi R_\mathrm{c}^3$ to mass and radius values from Figs. \ref{fig:mass-radius-relation}a) and b), the resulting mean density in the respective stellar is shown in  Fig. \ref{fig:mass-radius-relation}c)).   It can be seen that the mean density in low-mass stellar groups is not only not lower but indeed higher than in high-mass clusters. However, the question is whether these average densities matters here. They are all well below the threshold density of $n$ = 10$^4$/pc$^3$ for a sizeable effect on the discs.  Therefore, we look in Fig. \ref{fig:King} b) at the stellar mass distributions again, but this time with the half-mass radius chosen according to observed relation given by \mbox{eq. \ref{eq:mass-radius}}. It can be seen that for these observationally motivated cluster sizes, independent of their mass, all clusters have areas where the central density exceeds the threshold density of $n$ = 10$^4$/pc$^3$. Actually, in low-mass clusters, the relative area above this threshold is higher than in clusters of higher mass. Given the relatively small number of stars in low-mass clusters, the actual number of stars in this area is also small. However, the fraction of stars in the high-density area is higher in low-mass clusters than in high-mass clusters. \\

Besides, using the number density as a measure for a potential effect of the environment has its limitations, and its physical meaning is questionable, especially for low-mass clusters. What really matters is the probability for close interactions or the mean distance travelled to have such an interaction. A star will have to have crossed the cluster several times in low mass clusters before it has such close interaction. Nevertheless, motivated by the above estimates of the theoretical central stellar densities in the observed low-mass clusters, we performed detailed simulations of the encounter dynamics in such environments to obtain a better understanding of the flyby dynamics in low-mass clusters.  In the following section, we describe the numerical method, including the approximations made in this approach.\\

\section{Numerical Method}
\label{sec:method}

\subsection{Cluster simulations}

Our simulations start during the embedded phase assuming that star formation is completed.  After a time t\textsubscript{emb}, gas expulsion is taking place, resulting in cluster expansion. Also, a standard method in cluster dynamics simulations \citep[for example,][]{Lada_1984,Scally_n_Clarke_2001,Banerjee:2017}, it is in some sense a toy model as it avoids the complexity and numerical challenges the formation of massive clusters still faces. Thus the simulation sets only in when all stars are formed, and we take the mass-radius relations discussed in section \ref{sec:low_mass_cluster_properties}  at face value.
Naturally, any high-mass cluster must go through a phase where it starts initially with a small number of stars and gradually grows to its final size. However, such cluster formation simulations are still limited to clusters containing $\ll$ 1000 stars, at the least for the resolution required here. Therefore, we revert to the often used method of following the stellar dynamics in the fully formed clusters instead. The limitations of this approach are discussed in section \ref{sec:discussion}.

We perform simulations of the dynamics of clusters containing $N$ stars using the simulation code Nbody6++. In the simulations, a star is represented by a test particle with a given mass, position and velocity. The particles' positions are chosen so that the resulting stellar number density distribution obeys a King profiles \citep{King_1966}. The reason is that King profiles seem to present young clusters well \citep{Hillenbrand_1998,Harfst_2010,nurnberger_2002}; however, there is no general observational agreement whether W=7, W=8 or W=9 is the best choice. Here we mostly use a King profile with $W_0 > 9$, but also discuss in section \ref{sec:discussion} the role of the chosen profile for our results. 

The individual test particles are assigned masses following the initial mass function (IMF) by \citet{Kroupa_2001} 
\begin{equation}\xi(m) \propto\left\{\begin{array}{ll}
m^{-0.3} & 0.01 \leq m / \mathrm{M}_{\odot}<0.08 \\
m^{-1.3} & 0.08 \leq m / \mathrm{M}_{\odot}<0.5 \\
m^{-2.3} & 0.5 \leq m / \mathrm{M}_{\odot},
\end{array}\right.\end{equation}
with the lower mass limit set to $0.08M_{\odot}$ (hydrogen burning limit) and an upper mass limit of $150M_{\odot}$, which is the accepted upper mass limit \citep{Figer_2005,Weidner_2006}.

\begin{table}[t!]
\caption{Initial cluster parameters for the simulation campaign using mass-radius dependencies.}
\begin{tabular}{lcccccccc}
        \hline
        \hline
        \begin{tabular}[c]{@{}c@{}}Model\\ {} \end{tabular} &
        \begin{tabular}[c]{@{}c@{}}N\\ {} \end{tabular} & \begin{tabular}[c]{@{}c@{}}N\textsubscript{sim}\\ {} \end{tabular} &
        \begin{tabular}[c]{@{}c@{}}M\textsubscript{c}\\ {[}M$_{\odot}${]}\end{tabular} & \begin{tabular}[c]{@{}c@{}}R\textsubscript{hm}\\ {[}pc{]}\end{tabular} & \begin{tabular}[c]{@{}c@{}}M\textsubscript{t}\\ {[}M$_{\odot}${]}\end{tabular} &  \begin{tabular}[c]{@{}c@{}}t\textsubscript{emb}\\ {[}Myr{]}\end{tabular} & \begin{tabular}[c]{@{}c@{}}t\textsubscript{cross}\\ {[}Myr{]}\end{tabular} & \begin{tabular}[c]{@{}c@{}}t\textsubscript{relax}\\ {[}Myr{]}\end{tabular}\\
        \hline
        Set 1 & $\gamma=1.7$\\
        \hline
        P200& 200 &      1685 &     117.99 &  0.26 &    393.31 & 2.0 &    0.10 &     2.10 \\
        P1000& 1000 &     604 &     589.97 &  0.67 &   1966.57 & 2.0 &    0.18 &    14.20 \\
        P4000& 4000 &     147 &    2359.88 &  1.51 &   7866.27 & 2.0 &    0.31 &    78.36 \\
        P10000& 10000 &    95 &    5899.71 &  2.59 &  19665.68 & 2.0 &    0.44 &   247.35 \\ 
                   \hline
         Set 2 & $\gamma$ = 2.0\\
        \hline
        K200& 200 &      1484 &      117.99 &  0.32 &    393.31 & 2.0 &   0.14 &     2.90 \\
        K1000& 1000 &     492 &      589.97 &  0.70 &   1966.57 & 2.0 &   0.19 &    15.01 \\
        K4000& 4000 &      110 &     2359.88 &  1.35 &   7866.27 & 2.0 &   0.26 &    65.87 \\
        K10000& 10000 &    71 &     5899.71 &  2.09 &  19665.68 & 2.0&    0.32 &   178.68 \\ 
 \hline
         Set 3 & $R_c$ = 1.3 pc\\
        \hline
        V200 & 200 &       864 &       117.99 &   1.3 &    393.31 &  2.0 &  1.11 &    23.48 \\
        V1000 & 1000 &     167 &       589.97 &   1.3 &   1966.57 &  2.0 &  0.50 &    38.40 \\
        V4000& 4000 &       61 &      2359.88 &   1.3 &   7866.27 &  2.0 &  0.25 &    62.37 \\
        V10000& 10000 &     12 &      9439.53 &   1.3 &  31465.09 &  2.0 &  0.12 &   105.00 \\ 
    
               \end{tabular}
               	\vspace{0.5em}
	\tablecomments{Here  N denotes the number of cluster members, N\textsubscript{sim} the number of simulations for each cluster model, t\textsubscript{emb} the duration of the embedded phase, M\textsubscript{c} the stellar mass of the cluster, and half-mass radius r\textsubscript{hm}, M\textsubscript{t} the total cluster mass  (stars $+$ gas), t\textsubscript{cross} the cluster crossing time and t\textsubscript{relax} the relaxation time. Set 1 is based on the relation found in \cite{Pfalzner_2016}, set 2 for the relation derived in section 2 for the data by \cite{Kuhn_2019} and set 3 for a constant cluster size as modelled in \cite{Vincke_2016}.}
        \label{tab:cluster_parameters}
\end{table}

Initially, all cluster stars are single stars, meaning primordial binaries as excluded. This approximation is chosen for simplicity in the consecutive processing of the effect of flybys on disc size, where the inclusion of primordial binaries would significantly increase the complexity. The absence of primordial binaries might lead to underestimating ejections from the cluster due to strong few-body interactions \citep{Heggie_1975,Hills_1975}. For a more detailed discussion of omitting binaries in this type of study, see \citet{Olczak_2006}. Besides, binaries can accelerate a cluster's expansion process and reduce the encounter rate faster \citep{Kroupa_2001,Kaczmarek_2011}. 

Potentially existing initial mass segregation in the clusters was neglected. The occurrence of mass segregation within clusters is still under debate, 
whether it is primordial---found in the dense core phase of the cluster formation process \citep{Dib_2010,Dib_2019}---or if it is a result of the early dynamical evolution of the cluster \citep{Pelupessy_2012,Dominguez_2017}. 

The cluster members are given a velocity following a Maxwellian distribution involving the random sampling of Gaussian deviates for the individual velocity components. We assumed that the cluster is in virial equilibrium. During the embedded phase, the gas cloud is implemented as a background potential. The effect of gas expulsion is modelled as instantaneous as the timescales for gas expulsion is sufficiently short \citep{geyer_burkert_2001,Melioli_2006,Portegies_Zwart_2010}. 
The gas removal leaves the cluster in a super-virial state, and as a result, the cluster expands to return to a new virial equilibrium state. Stellar evolution has not been included in this work as it has limited influence on the results. Especially on the low-mass clusters that we focus on here. 

 We performed three sets of simulations differing in the choice of the cluster size. For Sets 1 and 3 the cluster half-mass radius was chosen according to eq. \ref{eq:mass-radius}, whereas for Set 2 the cluster size was a fixed value of 1.3 pc. For \mbox{Set 1}, $C$ and $\gamma$ where chosen according to \cite{Pfalzner_2016}, for Set 3 we used the values derived in Section \ref{sec:low_mass_cluster_properties} based on the data by \cite{Kuhn_2019}. For each of these sets, we performed simulations for clusters with $N$ = 200, 1000, 4000, and 10 000 test particles. For set 2 we modelled also clusters with $N$ = 16000 to be able to compare to the  results \cite{Pfalzner_2016} for high-mass clusters.  The simulation parameters are summarized in Table \ref{tab:cluster_parameters}. 
 It also provides the timescales relevant to cluster dynamics, namely, the crossing time, t\textsubscript{cross}, and the two-body relaxation time, t\textsubscript{relax}. Statistical significance of the results has been ensured by always performing many simulations ($N\textsubscript{sim}$) with different seeds for the spatial distribution.
This number of runs was chosen so that $N\textsubscript{stars} \times N\textsubscript{sim} \approx 5 \times 10^5$, which is necessary to obtain statistically significant results. More detailed information on the cluster dynamics method and its limitations is given in \cite{Vincke_2016}.

Ideally,  one would simulate the disc evolution around each star within the cluster-evolution simulation to determine the effect of the cluster environment on the discs. However, currently, such simulations are computationally too expensive to be performed for clusters with large $N$, unless one restricts them to a single simulation, simulation periods of \mbox{$\ll$ 1 Myr} and/or single-mass representation \citep{Rosotti_2014}. This latter approach does not allow the comparative statistically significant study we envisage here. Therefore, we apply a two-step approach, where the first step consists of the above-described cluster simulations. While running these simulations, the encounter history for each star is recorded, meaning for each star, the mass ratio and periastron distance of each close flyby is listed.  In the second step,  the truncation disc size for each of the close flyby events is calculated. Afterwards, we use the mean value of the final truncation disc size to measure the cluster environment's effect on forming planetary systems.

\subsection{Disc truncation}

We start with the artificial situation of all disc having a size of 1000 au initially. This value is typical for the top-end of the size of observed discs. \citep{Andrews:2018}. The reason is that we only intend to obtain a measure of the influence of the environment rather than a realistic representation of the disc sizes in a specific cluster. Thus we obtain the size that a standard disc is truncated to due to the close flybys of a particular cluster environment. This truncation disc size is obtained the following way: 

 In this work, all encounters were assumed to be parabolic, prograde and coplanar, using the results of a parameter study of star-disc encounters performed by \citet{Breslau_2014}. In this case, the truncation disc size $r_{disc}$ after the flyby can be approximated as
\begin{equation}
r_{\text {disc }}=\left\{\begin{array}{ll}
0.28 \cdot r_{\text {peri }} \cdot M_{21}^{-0.32}, & \text {if } r_{\text {disc }}<r_{\text {previous }} \\
r_{\text {previous }},                             & \text {if } r_{\text {disc }} \geq r_{\text {previous }}
\end{array}\right.
\label{eq:disc_size_est}
\end{equation}
where $r_{peri}$ is the periastron distance between the two stars involved in a close flyby in au. $M_{21}=M_{2} / M_{1}$ is the mass ratio of the two stars, with $M_1$ being the mass of the disc-hosting star and $M_2$ that of the perturber star. $r_{\text {previous }}$ is the disc size before the encounter. 

Inclined or retrograde flybys are less efficient to remove material from the disc, and the disc truncation radii are therefore larger \citep{Clarke_n_Pringle_1993,Bhandare_2016}. However, the difference to coplanar flybys is not extremely large. Applying the result for coplanar flybys means that the truncation radii have to be regarded as lower limits.
Similar, hyperbolic flybys are less effective in disc truncation than parabolic ones. However, more eccentric encounters are significantly more common in higher mass clusters. Since our study focuses on low-mass clusters, approximating the flyby path as parabolic is likely justified. Moreover, we apply these simplifications in all simulations identically, so the relative differences between the diverse cluster environments should not be affected. Here, only a summary of the method has been provided; we refer the reader to \cite{Breslau_2014} for more information on the disc size determination during close stellar encounters.

\subsection{Observational uncertainties}
\label{sec:discussion:uncertain}

As mentioned in section \ref{sec:intro}, the exponent in the derived mass-radius relations for young stellar clusters in the solar neighbourhood varies between $\gamma$= 1.7 --- 2.0  and that of the constant spans the range of $C$=  359 --- 1263. These variations are caused by the differences in the sample of clusters considered, the observation being at different wavelengths, and different cluster mass and size definitions.  In some studies, the cluster size is defined as the maximum extent of the cluster; in others, the half-mass or half-light radius. Similarly, the definition of cluster mass varies from study to study. Some studies give the number of observed stars, while some correct for the IMF. Therefore, one finds sometimes, even for the same clusters, differences of up to a factor 2. A detailed discussion of the involved difficulties can be found in \citep{Pfalzner_2016}.

As Fig. \ref{eq:mass-radius}c) shows these differences in mass and radius translates into equivalent differences in stellar density. Therefore one could worry whether the specific mass-radius relations studied here are of any relevance. In this study, we included two extreme cases in the sense that they cover the broad range of possible values for the mass-radius relation. Set 1 of our simulations (P200-P10000) stands for the strongest dependence on the cluster mass, whereas Set 2 covers the weakest case. In the following, we show the results from both sets, 1 and 2, and contrast them to those of set 3, standing for a constant cluster size to illustrate the extent to which previous simulations might have underestimated the effect of close stellar flybys in low-mass clusters.

The mass-radius relation is only well constrained for clusters with $N \leq$ 2000 stars as there are many clusters in the solar neighbourhood in this membership range.  Clusters containing $N \geq$ 2000 are rare in the solar neighbourhood. There are only a handful of such member-rich young clusters. Therefore, the mass-radius relationship is probably only applicable to clusters with $N<$ 5000 stars. We discuss this issue in more detail in section \ref{sec:limitations}.\\

\newpage
\section{Simulation results}
\label{sec:results}

\begin{figure*}[t]
\centering\includegraphics[width=0.80\textwidth]{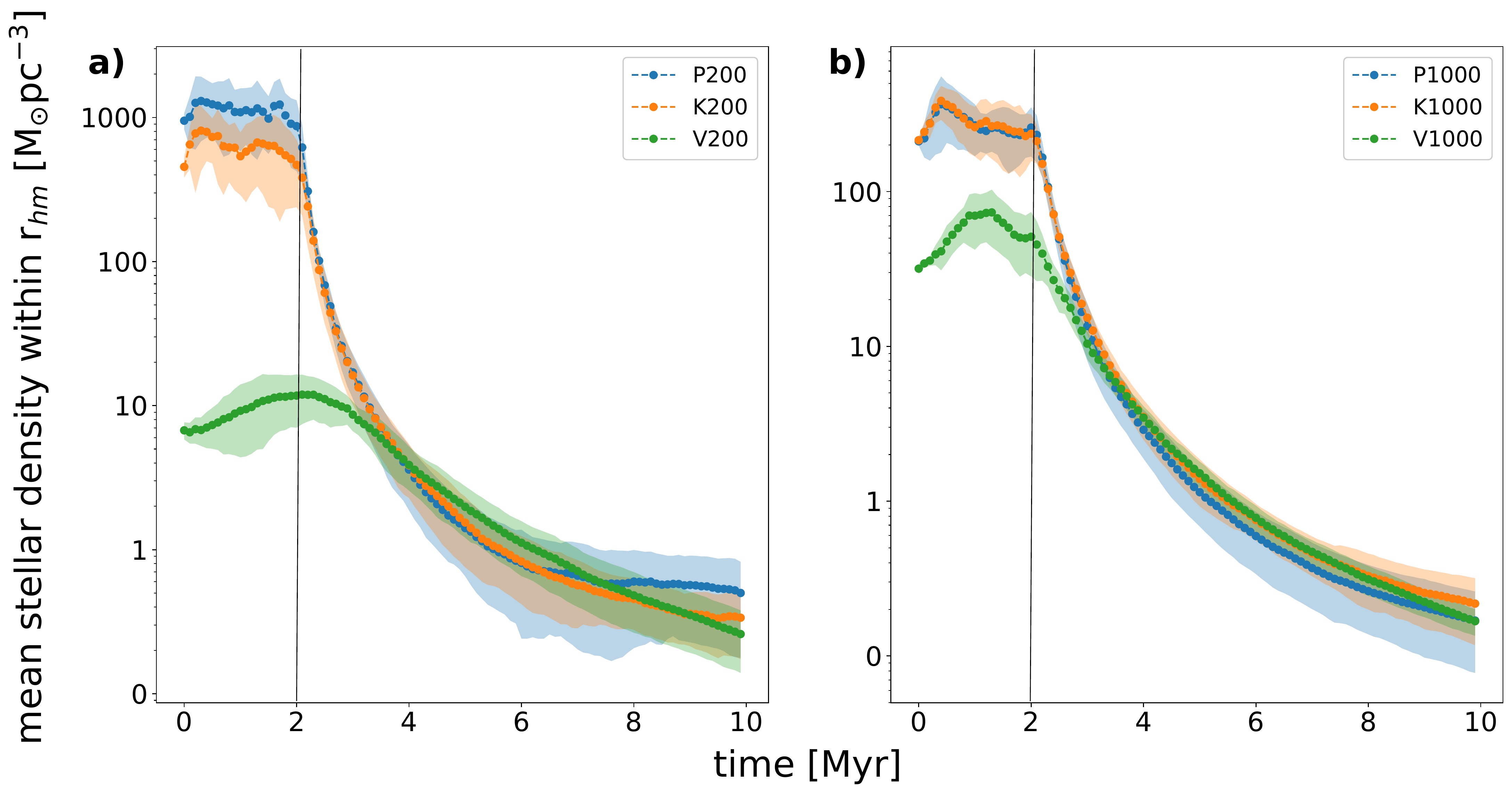}
\centering\includegraphics[width=0.80\textwidth]{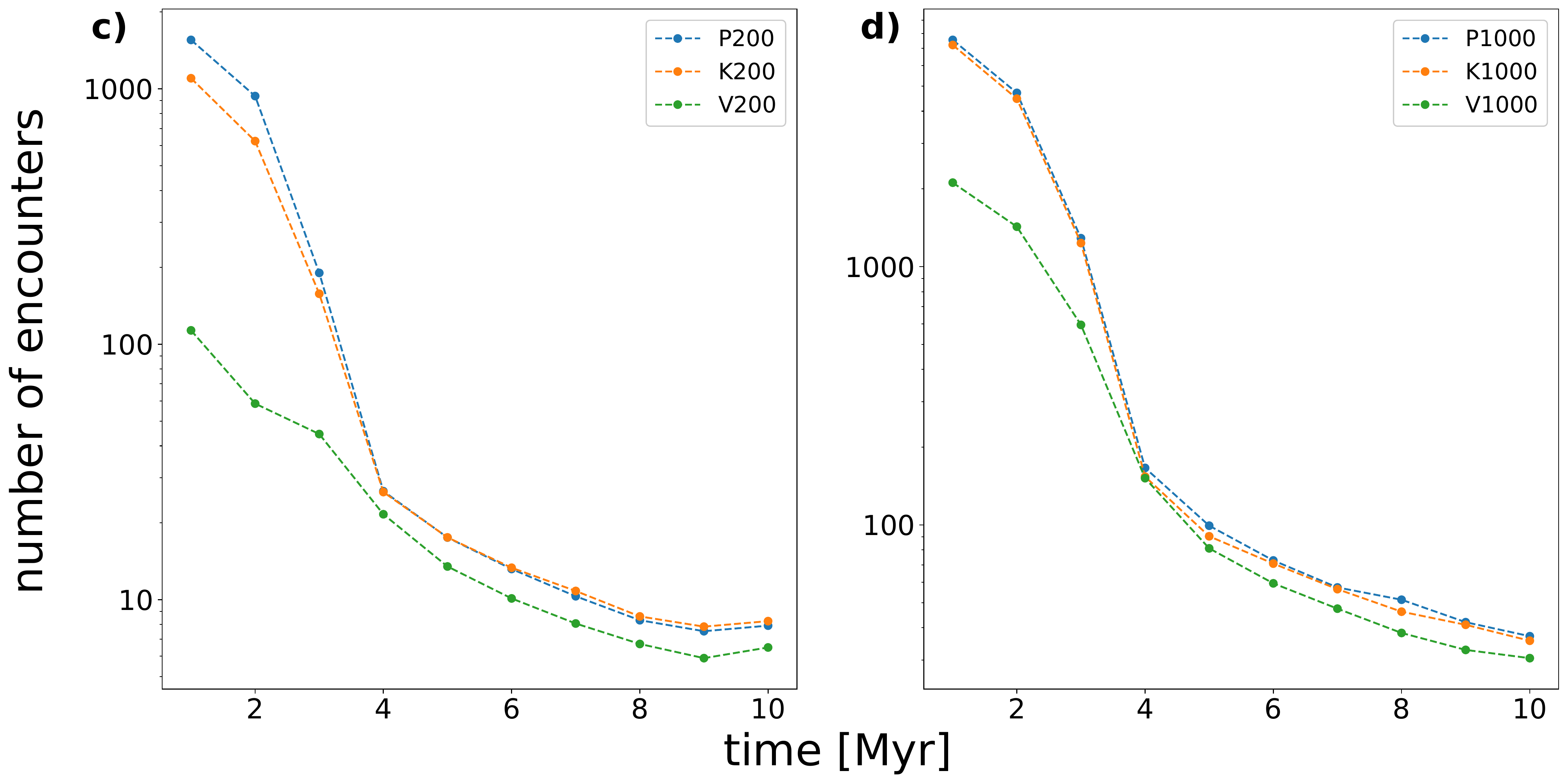}
\caption{Mean stellar density as a function of time for initial cluster parameters chosen according to the observed mass-radius relations ( blue --- $\gamma$ = 1.7 and green - $\gamma$ = 2.0) and a mass-independent size before gas expulsion. Here shown are the cases a) $N$ = 200 and b) $N$ = 1000. In all cases instantaneous gas expulsion takes place $t\textsubscript{emb}$ = 2 Myr indicated by the vertical black line.}
\label{fig:mass_density_evolution}
\end{figure*}

In section \ref{sec:low_mass_cluster_properties} we demonstrated that for the clusters in the solar neighbourhood initial conditions obeying a mass-radius (eq. \ref{eq:mass-radius}) are more realistic than those assuming a constant radius. Here we will investigate the consequences for the interaction dynamics in low-mass clusters. First, we study the development of the mean stellar density with time for low-mass clusters. The mean stellar density corresponds to the mass contained density in the volume defined by the cluster half-mass radius (equivalent to the 50\% Lagrangian radius).  Fig. \ref{fig:mass_density_evolution} shows that all low-mass clusters ($N \leq$ 1000) follow the same overall evolutionary pattern. 
During the initial embedded phase, the stellar density remains almost constant. However, when the gas is expelled, the clusters expand, which means a significant decrease in the mean stellar density. The cluster settles eventually in a new equilibrium state to a much lower density. Overall, the decline in stellar density can be quite dramatic.  Thus, our results confirm the qualitative picture of previous studies \citep{Lada_1984,Adams_2001,Kroupa_2001a,Boily_2003,Baumgardt_2007}.

The difference in the encounter history of clusters obeying the mass-radius relation lies on the quantitative level (see Fig. \ref{fig:mass_density_evolution}a). Whereas for the standard model V200 the mean stellar density is about 10 \MSun pc$^{-3}$ during the 2 Myr of the embedded phase, for K200 and P200 it is $\approx$ 1000  \MSun pc$^{-3}$, therefore, about a factor of a hundred higher for models P200 and K200. All clusters decline to approximately the same mean density of $\approx$ 0.5 --- 1.0  \MSun pc$^{-3}$,  after gas expulsion. However, the standard model only decreases to a tenth of its initial value, whereas models P 200 and K 200 experience a much more drastic decrease to a factor of $\approx$ 1000 lower mean stellar density. Besides, P 200 and K 200 have already reached their new (semi-) equilibrium state
after 6 Myr and 8 Myr respectively, whereas V200 keeps steadily declining as it has not reached its new equilibrium at the end of the simulation at 10 Myr and will probably require 20 --- 40 Myr to reach this state. 

Figure \ref{fig:mass_density_evolution}b) shows the same situation of clusters containing 1000 stars, where the difference between the various models is much less pronounced. In the embedded phase, model V1000 shows only a factor 7---10 lower mean stellar density than the two other sets. The mean stellar density in models P1000 and V1000 is nearly indistinguishable over the entire period of 10 Myr.  Our simulations of high-mass clusters $N$ = 4000 show that the differences in stellar density vanish completely. In summary, for clusters with $N<$ 1000, taking the mass-radius relation matters most.

Naturally, the encounter history directly reflects the development of the stellar density (see \ref{fig:mass_density_evolution}c and d). Here we defined encounters as those stellar flybys which have the potential to reduce the disc size to less than 100 au using the equation given in \citet{Breslau_2014} (eq. \ref{eq:disc_size_est}). The first few time steps are to some degree affected by the random setup. This leads to some artificial close encounters during the first few time steps, which becomes obvious in the low-number statistics of V200. Afterwards, the encounter frequency is about a factor 20 --- 25 higher for models P200 and K200, than for the standard model V200 during the embedded phase (see Fig. \ref{fig:mass_density_evolution}b)). The difference reduces to a factor of a few after ages 4 $>$ Myr.

\begin{figure}[!t]
	\centering\includegraphics[width=0.45\textwidth]{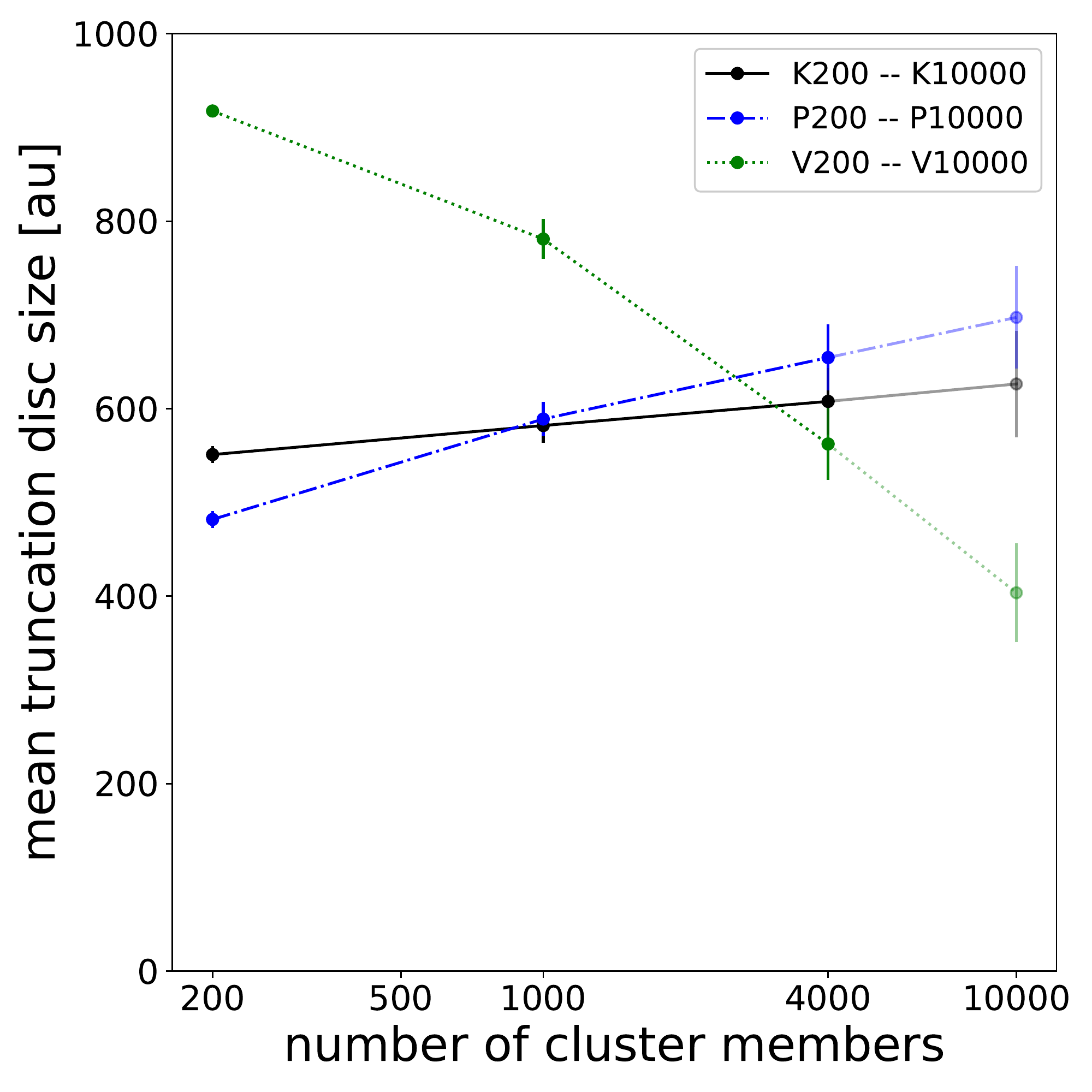}
    \caption{The mean truncation disc size as a function of the number of cluster stars. The two sets with mass-dependent initial are shown in blue and black, respectively. The set with a constant initial cluster radius is indicated in green. The mass-radius relation probably does to hold for clusters with $N>$ 5000, therefore the data points are shown in a lighter colour.}
    \label{fig:mean_disc_size_only}
\end{figure}

Next we want to define a measure for the effect of stellar flybys on discs and forming planetary systems. Here we use the mean truncation disc size, which we obtain by initially attributing a 1000 au-sized disc to each star. Using the recorded flyby history, we determine the truncation of the disc using eq. \ref{eq:disc_size_est} for each star due to all flybys that it encounters over 10 Myr. This truncation disc radius should not be confused with observed disc sizes, as most stars will have significantly smaller initial disc sizes than 1000 au. At this point, it is just a measure that allows the comparison of the strengths of the environmental effect in different cluster environments.
The relation to observed disc sizes is discussed in Section \ref{sec:discussion}.

From the just discussed encounter history, it seems already obvious that the diverse low-mass cluster models should lead to different truncation disc sizes. Figure \ref{fig:mean_disc_size_only} shows the mean disc size within clusters at an age of $10$ Myr as a function of number of cluster members for all cluster models listed in table \ref{tab:cluster_parameters}. To start with, we only consider clusters containing between 200 and 4000 stars.  The reason is that the mass-radius relation is only observationally constraint for this mass range and probably not applicable for clusters containing several ten thousand stars. For the low-mass clusters ($N$=200),  the mean truncation disc size is $\approx$ 900 au for the standard model (V 200). Only a few stars actually experience close flybys. By contrast, the observationally motivated models lead to much lower mean truncation radii --- 481 au for model P200 and 551 au for model K200. This shows that the effect of the environment in real embedded clusters is much stronger than mostly anticipated so far. Our results translate in a mean close flyby distance of 1000 au  -- 1500 au in low-mass clusters. This agrees with the results of $\approx$ 1000 au by \citet{Proszkow:2009} for their low-mass clusters ($N$ = 300). 

However, Fig. \ref{fig:mean_disc_size_only} contains an additional important information. Namely, that the overall effect of stellar flybys should be approximately the same for clusters that contain 200 stars as those containing 4000 stars. The latter is representative for clusters similar to the ONC. For model set 2 we obtain a mean disc truncation radius of 551 au for N=200 (K200) and 607 au for N=4000 (K4000). For model set 1, the lower truncation radius for low-mass relative to high-mass clusters would even indicated a stronger flyby dynamics in low mass clusters. This is in stark contrast to the standard model, where the mean truncation disc size for low-mass clusters is relatively large with $\overline{r}_d(V200) \approx $ 918 au compared to $\overline{r}_d(V10000) \approx $ 403 au for the high-mass clusters, which reproduces the results presented in \cite{Vincke_2016}. Their main finding was that given the same initial cluster radius, the median truncation disc size decreased for an increasing number of cluster members.

\begin{figure}[t]
\includegraphics[width=\textwidth]{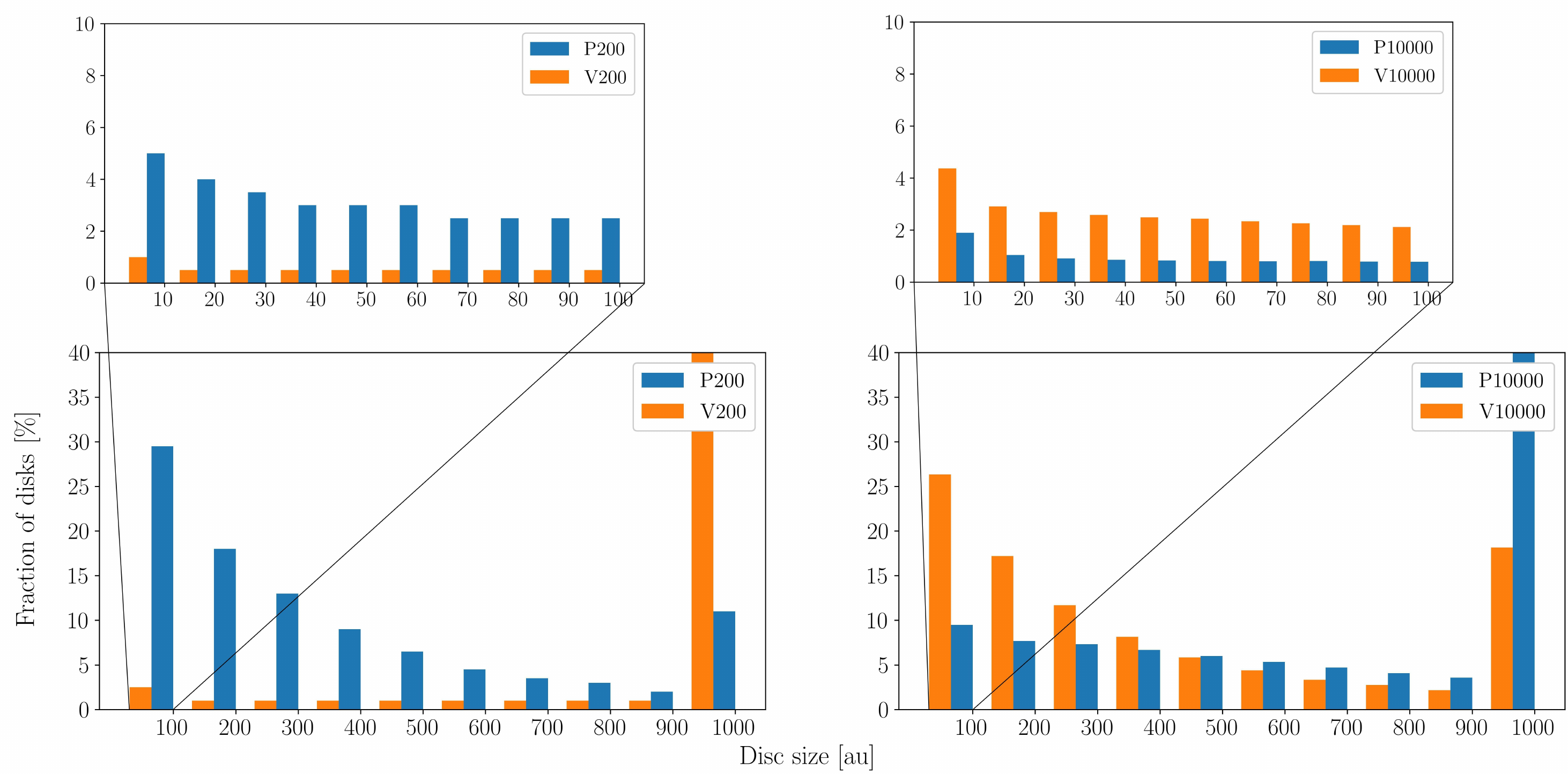}
\caption{Disc size distribution for in different cluster models. The left column shows the low-mass clusters models P200 and V200, the right column the high-mass clusters models P10000 and V10000. Set 1 is represented by blue bars, set 3 by orange bars. The bottom panels show the size range from 0 - 1000 au, whereas the top panel displays a zoom in on discs smaller than 100 au.}
\label{fig:dis_size_distribution}
\end{figure}

As discussed in section \ref{sec:discussion:uncertain} the mass-radius relation does probably not hold for clusters with more than  5000 members. Therefore, the disc truncation radii are shown in a lighter colour in Fig. \ref{fig:mean_disc_size_only}. We expect that for $N$=10 000, the standard model is a better approximation of reality. Here the disc truncation radius is slightly smaller (420 au) than for the low-mass clusters, and the effect of the flybys a bit stronger than in low-N clusters. Nevertheless, the difference in the effect of flybys in low-N and high-N clusters is much less pronounced than previously anticipated.

However, the mean truncation disc size gives only limited information about the interaction dynamics in clusters. The distribution of truncation disc sizes gives some valuable additional insights. The left panel of Fig.\ref{fig:dis_size_distribution} shows the distribution of truncation disc sizes for low-mass clusters (N=200), the right panel that for high-mass clusters {N=10 000}. In both panels, the results for an initial constant cluster size of 1.3 pc (orange bars) are compared to those obeying the mass-radius relation (blue bars). The distribution is shown over the entire size range from 0 --- 1000 au (bottom panels), but also a zoom in for disc sizes $<$ 100 au is given with bin sizes of 10 au (top panel). 
Looking at the lower panel, one sees that the truncation disc size distributions for models P200 and V200 differ considerably. In the standard model V200 (orange bars), approximately 95\% of all discs have a disc size close to the initial disc size of 1000 au. Those are virtually unaffected by close flybys. In model P200, the situation is very different, with nearly 87\% of discs having disc sizes much small than 1000 au at the end of the simulation. This high percentage for model P 200 means that most discs are affected by close flybys, with close to 30\% being reduced to sizes smaller than 100 au. Zooming in on the discs with sizes less than 100 in the top panel of Fig.\ref{fig:dis_size_distribution} astonishingly, in model P 200, approximately 5\% of discs are even truncated to sizes of less than 10 au. This result strongly indicates that the frequency of close stellar flybys has been severely underestimated for low-mass clusters. 

The truncation disc size distribution in model P200 (blue bars in the left-hand panel) is surprisingly similar to that of model V10000, the standard model for high-N clusters (orange bars in the right-hand panel). In P200,  80\% of discs are significantly affected by close flybys.  Approximately 27\% of discs are even reduced to less than 100 au, and approximately 4.5\% of discs truncated to less than 10 au. The results are very similar for the steeper mass-radius relation derived from the Kuhn data (as shown in Fig. \ref{fig:dis_size_distribution_k200}). Considering these fractions of truncated discs demands a fundamental change of our view of the role of the environment in low-mass clusters: Namely, the effect of the environment by close stellar flybys is approximately the same in high- and low-mass short-lived clusters. 

Our results demand a considerable change in how we think about the influence of the environment in low-mass clusters. Therefore, we want to make sure that the result holds against closer scrutiny.
Therefore, we performed some additional tests to exclude that particular choices in our simulations lead to an overabundance of close flybys in low-mass clusters.

\begin{figure}[t]
\includegraphics[width=0.446\textwidth]{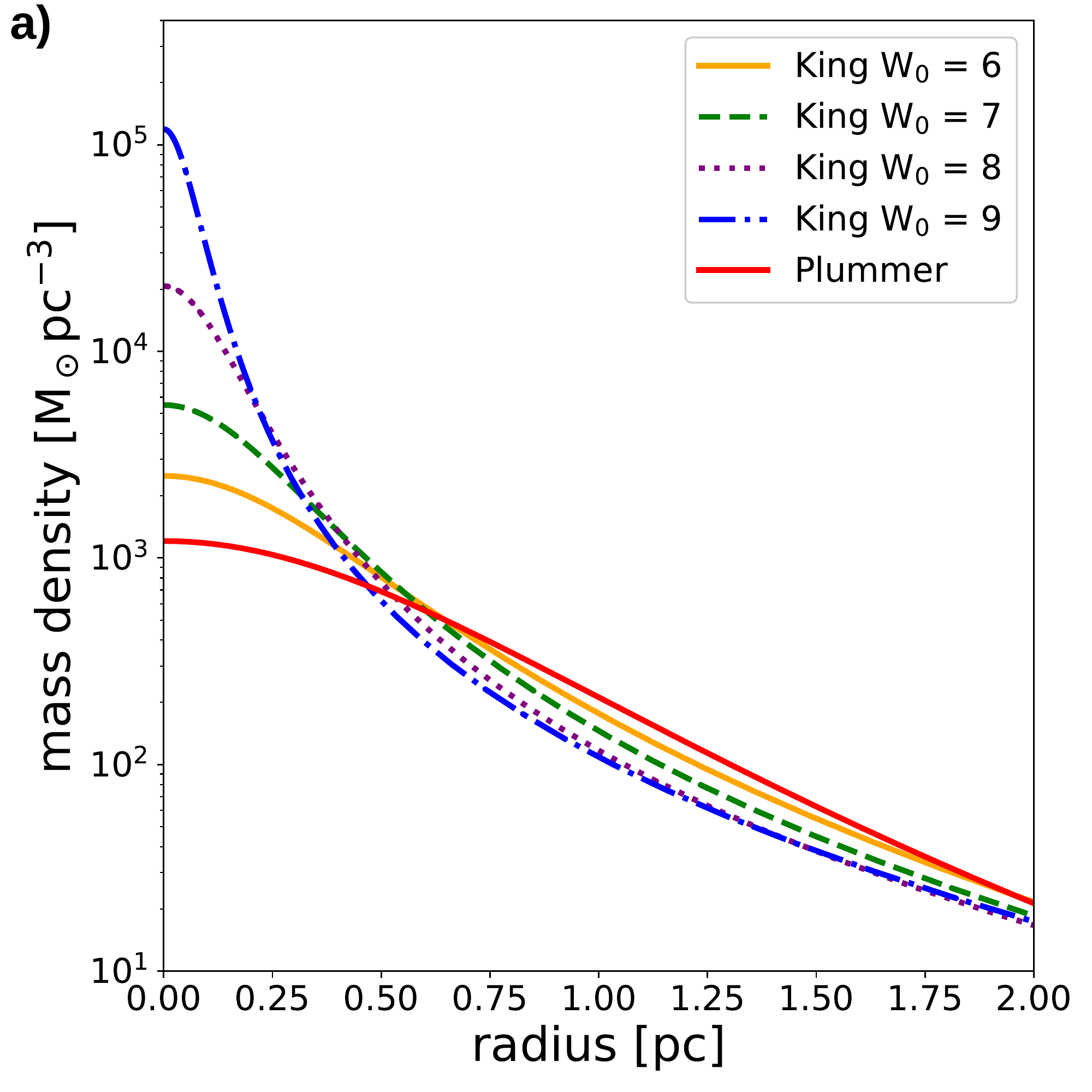}
\includegraphics[width=0.45\textwidth]{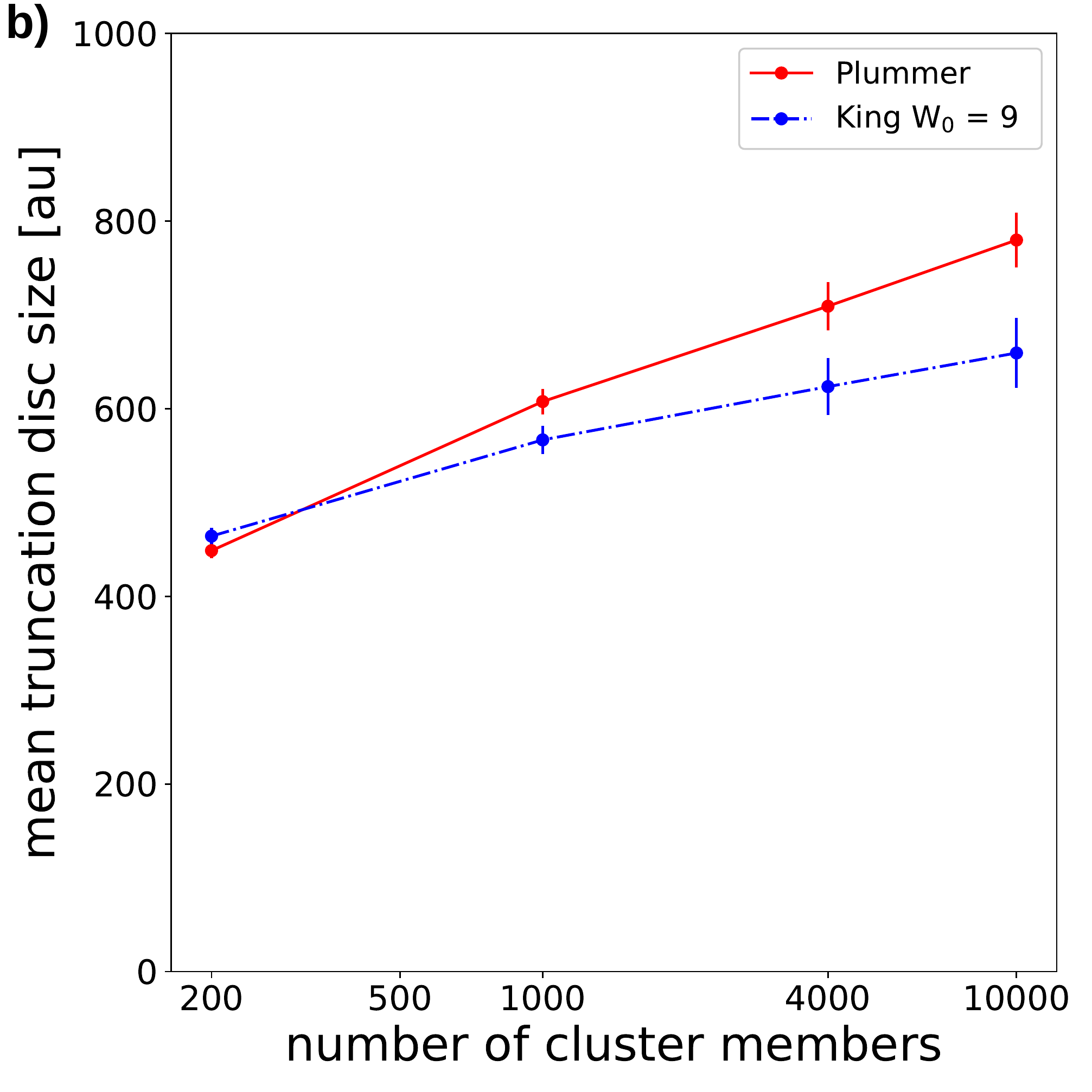}
\caption{a) Comparison of mass density as a function of distance to cluster centre for Plummer and King profiles for a cluster with a mass of M= 5000 M$_{\odot}$ and a half-mass radius of 1.3 pc. b) same as Figure \ref{fig:mean_disc_size_only}, but here a comparison for model P 200 assuming a Plummer  and a King (W=9) profile is shown. 
}
\label{fig:density_profile}
\end{figure}

In section \ref{sec:method} we chose a King W = 9 profile for the stellar distribution, as this profile corresponds to the observed density profile of the ONC \citep{Hillenbrand_1998}. However, low-mass clusters contain fewer stars, so that profile fitting is less well constrained than for high-mass clusters. Therefore, it is unclear whether a King (W=7) or the even much flatter Plummer profile are equally good or even better choices for low-mass clusters. Fig. \ref{fig:density_profile}a shows a comparison of different King profiles (W = 6, 7, 8, and 9) and a Plummer profile for a cluster of mass 5000 M$_{\odot}$ and a half-mass radius of 1.3 pc. The central stellar density is highest for a King profile W=9 and lowest for a Plummer profile; this leads to more close flybys than in clusters with a Plummer profile. Therefore, there exists the potential risk that the choice of profile could give unrealistically high rates of close encounters in low-mass clusters. 
We tested whether the choice of the stellar density profile is responsible for the high frequency of close flybys in low-mass clusters by also performing simulations with a Plummer profile for the entire set of cluster parameters. Figure \ref{fig:density_profile}b) shows a comparison of the mean truncation disc size shown in Fig.\ref{fig:mean_disc_size_only} obtained by using a King profile with those applying a Plummer profile. It can be seen that, for high-mass clusters ( N = 10 000 ) in King-type clusters, discs are on average truncated to smaller sizes than in Plummer type clusters. This result confirms the expectation. However, there is virtually no difference between the mean truncation radius for the two types of profiles for low-mass clusters. The reason is that in high-mass clusters, a considerable amount of stars actually stay for a relatively long time in the central high-density. In contrast, in low-mass clusters, the stars quickly pass through this area. In conclusion, our result for the low-mass clusters still holds if we apply a Plummer distribution instead of a King W=9 profile.


As shown in table \ref{tab:cluster_parameters} we simulated 1685 different realisations of model P200. The stellar masses are randomly chosen according to the IMF; therefore, some low-mass clusters contain high-mass stars. Such high-mass stars inevitably lead to gravitational focusing, which can cause rapid disc destruction. Next, we test whether the result could be biased by gravitational focusing in those low-mass clusters containing high-mass stars. Therefore, we divided our complete set of simulations into two subsets. One consisted of all those clusters that contained one or more high-mass star ($M_s >$ 20 \MSun), and another subset that was devoid of high-mass stars. We obtained approximately the same values for the disc truncation radius for both subsets and the entire ensemble. We conclude that clusters containing high-mass stars do not influence the result in any way.

\section{Consequences and observational tests}
\label{sec:discussion}
From our simulations, it follows that in low-mass clusters, the effect of the environment on discs is comparable to that in high-mass clusters. The question is how one can test this hypothesis.  In the following, we point out possibilities how high-resolution observations of discs in low-mass clusters and statistics of exoplanet orbital properties could be utilized to test this hypothesis.  

\subsection{Disc sizes and outer edge geometry}

We pointed out in section 2 that we start our simulations with all-stars being initially surrounded by a  1000 au-sized disc. Therefore, the resulting disc sizes are referred to as truncation disc sizes as they are primarily a measure of the interaction dynamics. In reality, clusters will likely form with a broad spectrum of disc sizes; however, the primordial disc size distribution is difficult to access, as it is determined in the observationally difficult to access deeply embedded phase. Most likely, most discs will be considerably smaller than the here used 1000 au. However, a comparison with observations might be feasible nevertheless for those discs that are truncated to sizes considerably smaller than 100 au. For them, the connection to the initial assumed disc size is erased by the close flyby. In the following, we will compare the relative number of discs smaller than 100 au in our simulation to those in observations being fully aware of the limitations of this approach.

\begin{figure}[t]
\centering\includegraphics[width=0.446\textwidth]{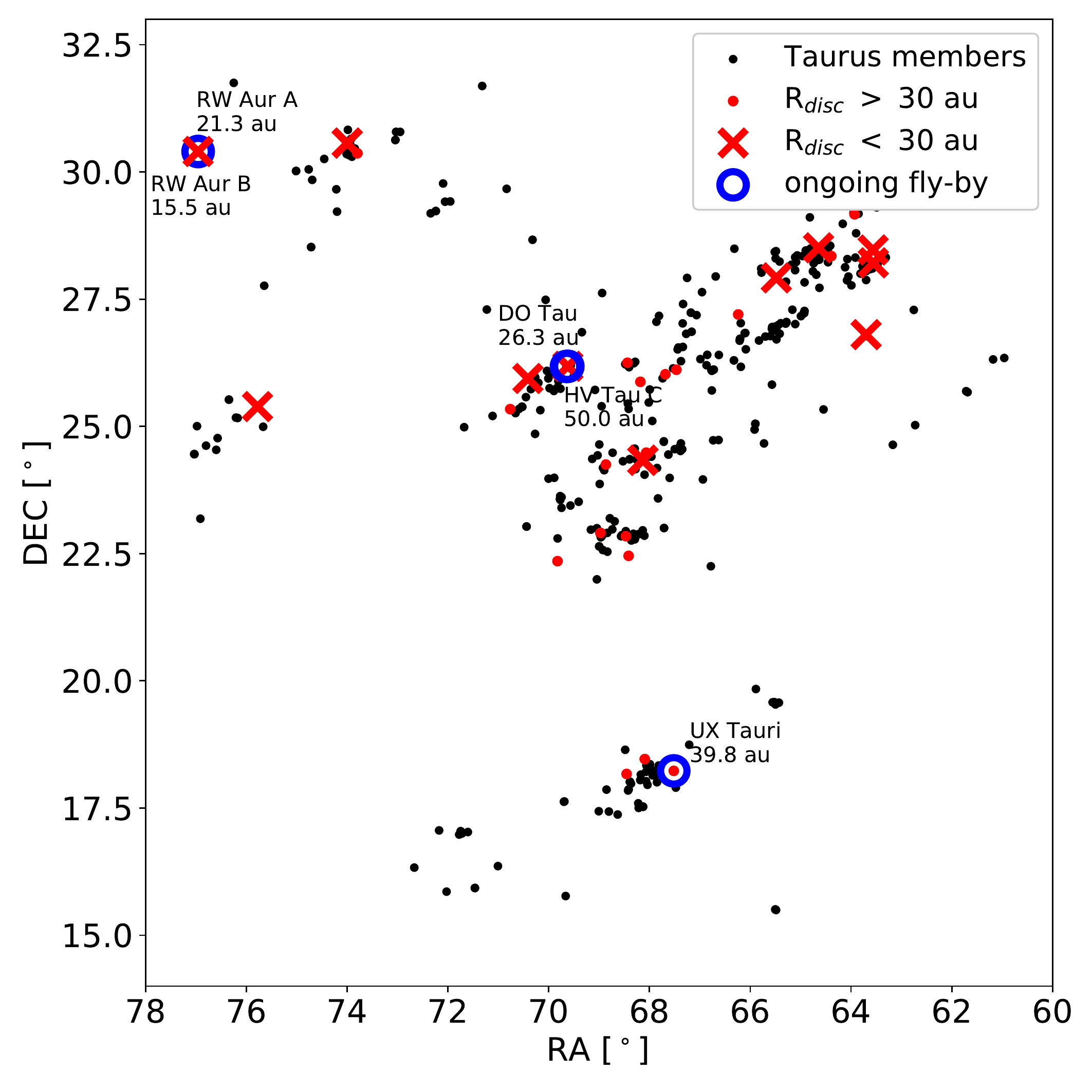}
\caption{ Location of the members of the Taurus star forming region according to \citet{Luhman_2018}. The systems showing spiral arms attributed to stellar flybys \citep[][]{Cuello:2019, Menard:2020} are indicated as blue circles. Note, that some are located close to each other, so that they are better identified by their labels. The red symbols indicate the location of the stars for which \citet{Andrews_2018} determined disc sizes. Discs with sizes smaller than 30 au are indicated as red crosses, whereas discs with sizes exceeding 30 au are shown as circles. The disc sizes for RW Aur A \& B were taken from \citet{Rodriguez_2018}}. 
\label{fig:taurus_discs}
\end{figure}

This work focuses on young low-mass clusters in the solar neighbourhood. An ideal example clusters to test our results would be IC 348 or LkH$\alpha$ 101. Unfortunately, for both clusters, no information about their disc size distribution exits so far. More information on disc sizes in the literature can be found for low-mass star-forming regions like Taurus-Auriga, Ophiuchus and Lupus. These regions do show clustering of stars; however, whether these stellar groups are actually bound and dynamically act as a cluster is uncertain. \citet{Luhman_2018} found a total of 427 candidate members of the Taurus region, which he divided into four groups with around 50--150 stars. Despite not being the ideal candidate, we concentrate our comparison on the Taurus region because five out of the eight examples where stellar flybys are suspect as the reason for the spiral arm structures are located in the Taurus region. Figure \ref{fig:taurus_discs} shows the location of these systems  (blue circles) in the Taurus region. While some system in isolation exist, the majority resides in regions of relatively high stellar density.

In the previous sections, we concentrated on modelling disc sizes. During the last few years, high-resolution observations of protoplanetary discs found that disc sizes span over a wide range, and that gas and dust disc sizes can differ considerably, with gas disc sizes usually being considerably larger than the dust disc sizes.  \citet{Tripathi_2017} find in their high-resolution (sub) millimetre survey of dust continuum emission in Taurus and Ophiuchus that in the  50 studied protoplanetary discs, the most common disc size are larger than 100 au, but some discs being considerably smaller. \citet{Andrews_2018} extended this study to include 56 discs in the Lupus SFR, finding a similar range in disc size as on Taurus but with relatively more small discs ($<$ 30 au). \citet{Nijita_2018} find that in the Taurus region, many discs are larger than 200 au, but there is a large spread in the sizes of Class II gas disks at any age, including a population of tiny Class II gas disks. They suggest that the small sizes may result from processes such as photoevaporation, disk winds, or truncation by orbiting low-mass companions.

The results presented here show that stellar interactions have the potential to produce such small discs even in low-mass clusters.
For our model P200, representative for low-mass clusters, we find that while
most discs 
are larger than 100 au, approximately 10\%--30\% of discs are considerably smaller than 100 au. This compares well with the observed fraction of small-sized discs in low-mass star forming regions. Figure \ref{fig:taurus_discs} also indicates the stars for which disc sizes have been measured (red symbols), where the dots indicate disc sizes $>$ 30 au  and crosses discs with sizes $<$ 30 au \citep{Andrews:2018}. It can be seen that most of these small discs are located in the relatively populated areas in Taurus, which agrees well with the expectations from our simulations. Possibly, close stellar flybys should not be excluded of hand as a cause for small discs in low-mass clusters. 

\begin{figure}[t]
\includegraphics[width=1.00\textwidth]{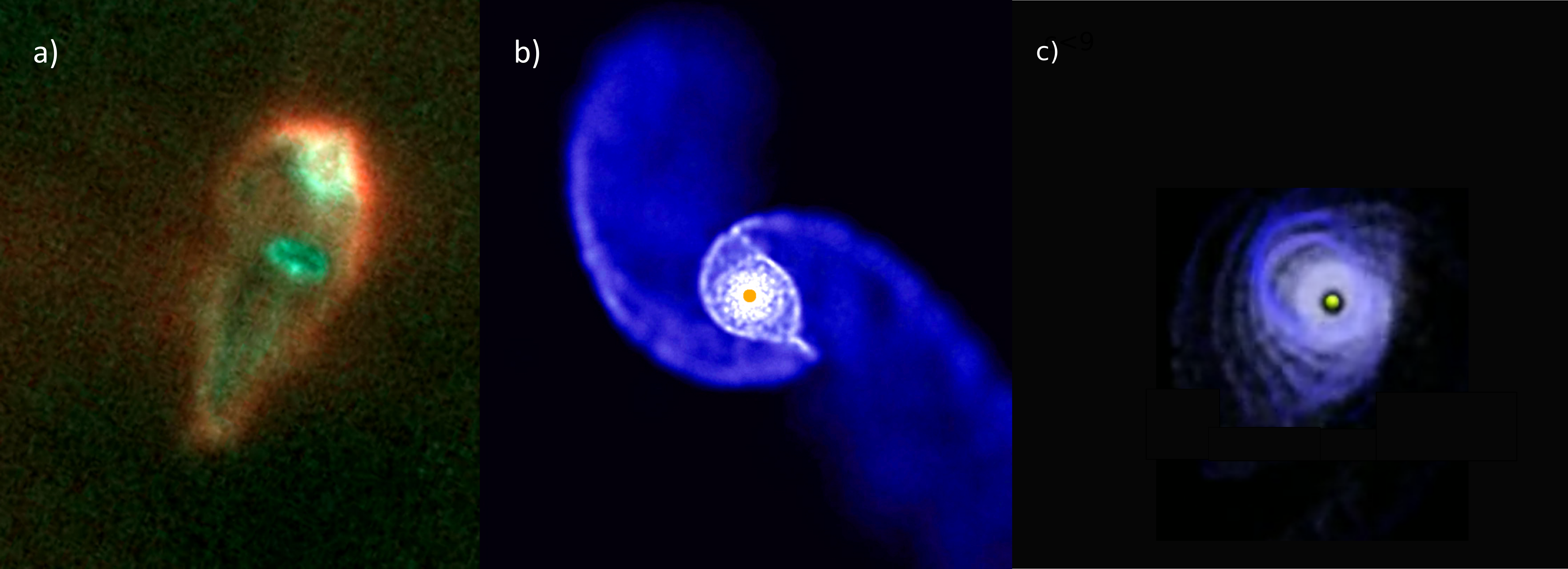}
\caption{Environmental effects on protoplanetary discs.
a) HST image of an externally photo-evaporated disc (proplyd) (credit: NASA/ESA). The  simulation results of the effect of a stellar flyby a) shortly after the encounter and b) 2000 years later.}
\label{fig:disc_shape}
\end{figure}

In section \ref{sec:results} we saw that close stellar flyby should lead to $\approx$ 13\% of discs having sizes of $\leq$ 30 au. However, testing for the portion of small disc sizes alone is not sufficient as these discs could equally just have formed small. However, there is a way to distinguish the two reasons for small discs.  The relevant property is the disc's outer edge. Discs that formed small should show a smooth decline in surface density all the way out. By contrast, truncated discs are distinguishable by their sharp outer edge (see Fig. \ref{fig:disc_shape}c).  This distinct edge will eventually be smeared out due to viscous spreading. However, it should be visible for at least 1-2 Myr. Therefore, the sharp outer edge should be most pronounced in relatively young clusters ($<$ 2 Myr), where viscous spreading has not had sufficient time to erase the sharp edge over this period.

External photo-evaporation can also lead to disc truncation. The question arises whether small discs caused by external photo-evaporation could be mistaken for those caused by close stellar flybys. Distinguishing small discs caused by flybys from those caused by external photo-evaporation likely is not a major problem for two reasons. First, most low-mass clusters contain no O- or B-stars that could cause external-photo-evaporation. Second, the shapes of small discs produced by flybys and external photo-evaporation differ considerably in their visual appearance. The best example for the discs that are affected by external photo-evaporation are the proplyds in Orion \citep{Ricci:2008} with their tadpole shape (see Fig. \ref{fig:disc_shape}a).). By contrast,
flybys lead to the characteristic spiral arms see Fig. \ref{fig:disc_shape}b) that quickly wind up, leaving behind a circular disc with a tightly wound ring structure (see Fig. \ref{fig:disc_shape}c). Thus mistaking one for the other is quite unlikely. 

The question arises why the influence of the environment in low-mass clusters has not become observationally apparent so far. Moreover,
there are several studies showing that disc masses or frequencies in the centres of high-mass clusters like the ONC are lower than in the same cluster or in low-mass clusters \citep[for a recent example, see ][]{Terwisga:2019}, which is interpreted as the presence of environmental effects in these regions. However, using the evidence of environmental effects in high-mass clusters to conclude that environmental effects play no role in low-mass clusters, because so far no evidence of it has been found,  would be a logical mistake.
There are several reasons why so far the influence of the environment in low-mass clusters might not have become observationally apparent yet.  First, there have been many dedicated observation campaigns to detect the effect of the environment in high-mass clusters, but none in low-mass clusters.  When one does not expect an effect, the motivations to search for it is low. Second, in low-mass clusters, one always deals with small-number statistics. Let us assume that in ONC-like clusters, 5\% of stars would be affected by the environment: This would still be a detectable effect, as it would concern 200 stars. For example, the prediction that the disc fraction or mass in the central area or close to the O stars is lower than elsewhere is relatively easy to test under these conditions. The situation is quite different for low-mass clusters where a 5\% effect would concern 10 stars. Third, and perhaps most relevant, the cluster crossing time in the low-mass clusters investigated here is of the order of 0.1 Myr. Thus in contrast to the situation in the high-mass clusters, these 10 affected stars would not be concentrated in the dense cluster centre but could be anywhere within the cluster. Therefore, one would have to compare the disc masses in the entire area of high-mass cluster to that in low-mass clusters to determine the relative importance of environmental effects in dependence of cluster mass.

\subsection{Planetary systems}

Close stellar flybys could affect resulting planetary disc structures in a variety of ways. What follows most directly from sections 4 and 5 is that small truncated discs hinder the formation of planets on wide orbits. Especially planets on orbits that exceed 100 au or 200 au should be less common than on smaller orbits. Unfortunately, currently, there are still far too few known exoplanets on distant orbits to test this effect, as current exoplanet detection methods favour the detection of close-in planets. It is an open question whether most stars are formed in low- or high-mass clusters. However, as our model predicts that there is virtually no difference in the influence of low- and high-mass clusters on forming planetary systems, this result would hold independent of the outcome of this discussion.

The time scale of planet formation relative to cluster lifetime is still an open question. If planets do indeed form within 1-2 Myr, close flybys could influence directly the structure of the planetary system \citep{Pfalzner:2018,DeRosa:2019,Winter:2020}. In this case, close stellar flybys primarily affect the outermost planets that form at the disc's edge. Close stellar flybys during that stage would often excite those outer planets onto strongly eccentric orbits. A preference for outer planets showing eccentric orbits would be another indication of the relevance of close flybys.\\

A typical effect of a close flyby is the resulting spiral arm structure that quickly develops into eccentric rings which circularize on timescales of a few 10 000 years. Another sign for close stellar flybys in low-mass clusters would be a higher rate of structured discs among the small discs than the more extended discs in low-mass clusters. Close flybys could promote or accelerate planet formation in varies ways also in small clusters. As flybys lead to the accumulation of material in these ring structures, this leads to localized higher dust densities, which is beneficial for dust growth. Similarly, close flybys can lead to gravitational instabilities that can trigger planet formation.

\subsection{ Cluster dynamics}

There is also a more indirect test of the above hypothesis. Low-mass clusters of the same size as high-mass clusters have a relaxation time of several tens of Myr after gas expulsion\citep[see for example,][]{Bastian:2006}. Thus these clusters would be well out of virial equilibrium at the age of 10 Myr.  However, again this assumes low- and high-mass clusters are of the same size. As observations tell us that this is not valid, and low-mass clusters are considerably smaller than high-mass clusters, low-mass clusters should already have reached virial equilibrium at ages of $\approx$ 5 Myr.  Therefore,  determining the velocity dispersion of low-mass clusters in the age range 5 --- 15 Myr would allow testing the above assumptions. If the above model holds, low-mass clusters should already be close to virial equilibrium at 10 Myr.  \\

\section{From low- to high-mass clusters}
\label{sec:limitations}

In section \ref{sec:results} we applied the commonly-used method of studying the dynamics of young clusters starting out with a fully-formed cluster. As mentioned in the introduction, this is done due to the prohibitive computational expense of modelling the cluster formation process itself for a statistically significant number of cases.
As pointed out, in this sense, our model has to be regarded as a toy model that does not account for the cluster formation process. Nevertheless, we try in the following to draw some conclusions concerning the close flyby history during the star formation process
in particular for high-mass clusters.

Any cluster starts out with a small number of stars that gradually increases until the star formation process ceases. Meaning any high-mass cluster was once a low-mass cluster. The problem is that in the formation phase, the clusters are still embedded in their gas. Therefore, their properties in terms of cluster mass and size are observationally much more difficult to constrain. One could consider two scenarios for high-mass cluster formation, which might lead to different close-flyby histories:

First, high-mass clusters could already form spread out about a more extensive area from the start and only become identified as stellar overdensities when they have already reached a higher mass. In this case, their close encounter rate would gradually increase as the stellar density increases. Second, high-mass clusters could start small and become more extended later on, following the mass-radius relation.  Then high-mass clusters would undergo an early phase where they experience the same close stellar flyby history as low-mass clusters during later stages in their life. The only difference would be that the amount of gas would be higher. This second scenario might be the more realistic one, as the mass-radius relation seems to be obeyed independent of the gas mass of the cluster. 

Although the close flyby history differs in these two scenarios,  the frequency of close encounters and the overall effect on the discs might be very similar. The difference lies more in the question of timing than inefficiency. In the first model, the frequency would steadily increase, be highest just before gas expulsion, and drop dramatically afterwards. In the second model, the low- and high-mass clusters have a surprisingly similar frequency of close encounters. The close encounter rate would be more or less constant until gas expulsion. More observations of clusters in the deeply embedded phase are necessary to distinguish between the two options.\\

\newpage
\section{Summary and Conclusion}

This study tests the hypothesis that close stellar flybys might be more common in young low-mass short-lived clusters ($N\approx$ 200) of the solar neighbourhood than anticipated so far. Our approach is based on the observational correlation between an embedded cluster's mass and size in the solar neighbourhood's stellar groups. We showed that the smaller size of low-mass clusters translates into a somewhat higher mean density than in high-mass clusters. Moreover, in low-mass clusters, not only the mean density is high, but the central stellar densities reach values above the threshold ($\rho$ = 10$^4$/pc$^3$) for close stellar flybys affecting forming planetary systems. Motivated by our findings, we performed simulations of the dynamics in low- and high-mass clusters obeying the mass-size relation. Recorded the close flyby history over the first 10 Myr of cluster development, and we find that

\begin{itemize}
\item in low-mass clusters, close flybys seem to be as common as in high-mass clusters like the ONC. The mean occurrence rate and periastron distance of close flybys are similar in both environments.  
\item In low-mass clusters, approximately 30\% of the discs are truncated to 100 au or less in size. This disc truncation happens predominantly in the central areas of these clusters.
\item Amazing 13\% of stars have flybys that lead to the truncation of their disc to less than 30 AU. Thus, even in low-mass clusters, 13\% disc would be too small to form equivalents of the solar system because they would be either stripped of their outer planets or hindered from forming them in the first place.  
\item Even in low-mass clusters, large discs with sizes of a thousand au or more have difficulties surviving.  Half the stars in low-mass clusters have a flyby closer than 1500 au, corresponding to a truncation disc size of approx 500 au for an equal mass encounter. 
\end{itemize}

Thus, the cluster environment's influence is relatively independent of the cluster's mass or, equivalently, the number of stars in it.  Only in clusters N$>$ 10,000  or compact long-lived clusters predominantly found in the spiral arms or the galactic centre, close flybys are considerably more common than in the low-mass clusters investigated here.

If confirmed, these findings would require a paradigm shift  in terms of the expectation of the relevance of close stellar flybys in low-mass clusters. So far, the expectation was that close stellar flybys are so rare in low-mass clusters that their effect on forming planetary systems is negligible. As a consequence, observations of low-mass clusters are nearly never checked for environmental influence. However,  our results suggest that low-mass clusters influence forming planetary system to the same degree as the ONC. 

These theoretical results lead to three observationally testable predictions. First, as we saw, close stellar flybys should lead to $\approx$ 13\% of discs having sizes of $\leq$ 30 au. However, testing for the portion of small disc sizes alone is not sufficient as these discs could equally just have formed small. The relevant property is the disc's outer edge, which should be sharp in contrast to shallow outer edges of the discs that formed small.  This sharp outer edge should be most pronounced in young clusters ($<$ 2 Myr), as viscous spreading has not had sufficient time to erase the sharp edge over this period. Second, a preference of outermost planets on wide orbits ($>$ 20 au) showing high eccentric would be another indication of the relevance of close flybys even in low-mass clusters. Third, 10---20 Myr-old low-mass clusters should be close to virial equilibrium.

\begin{acknowledgements}
We thank the referee, who has provided very helpful comments which have improved the manuscript.
\end{acknowledgements}


\appendix

\begin{figure}[!h]
\centering\includegraphics[width=8.6cm]{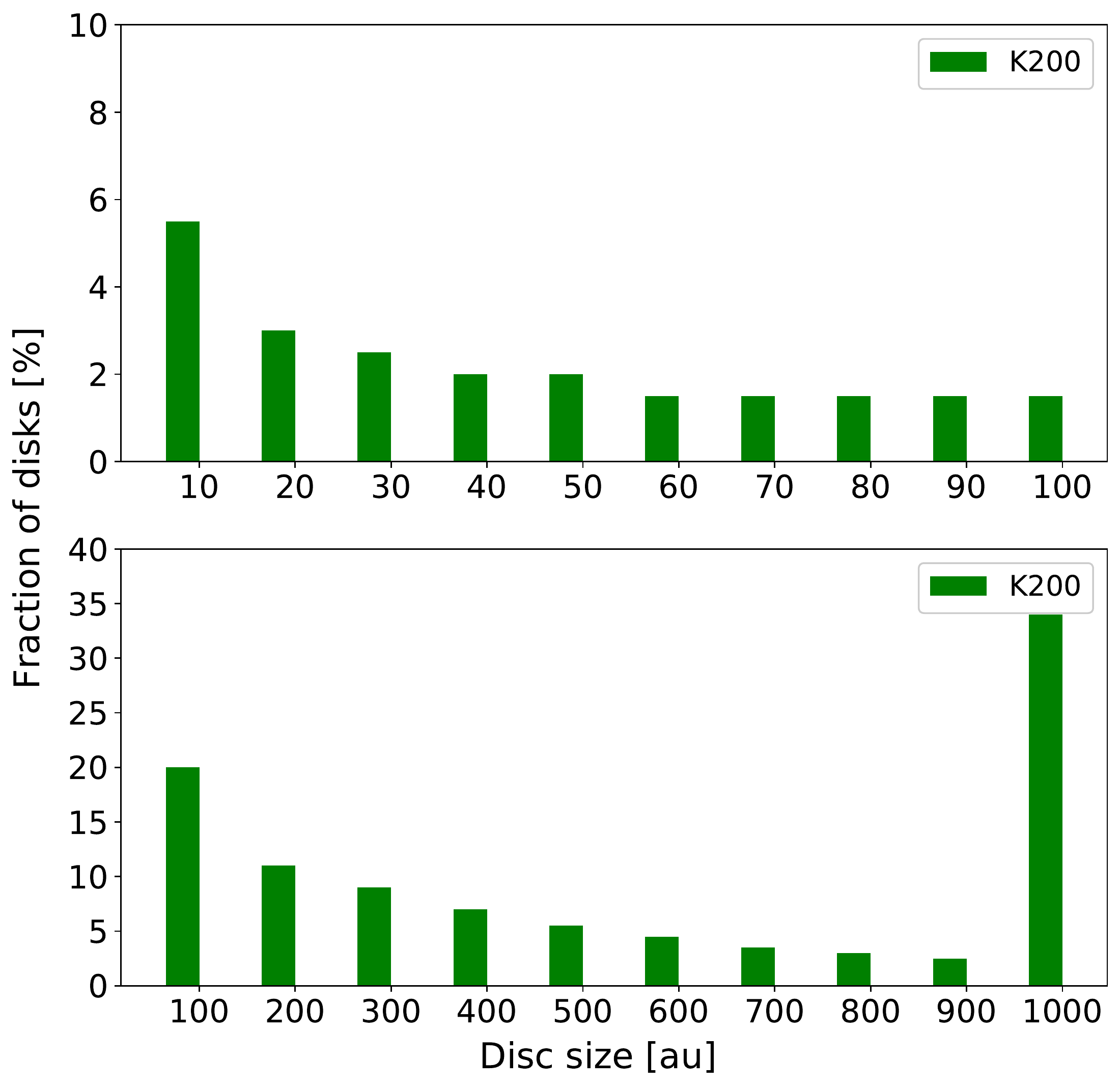}

\caption{Disc size distribution for cluster K200. The bottom panels show the size range from 0 - 1000 au, whereas the top panel displays a zoom in on discs smaller than 100 au.}
\label{fig:dis_size_distribution_k200}
\end{figure}

In Section \ref{sec:results} we have shown that taking the mass-radius relation into account low-mass cluster have a small disc size as in the standard model V200. Figure \ref{fig:dis_size_distribution_k200} shows the disc size distribution for model V200 for comparison with Fig. \ref{fig:dis_size_distribution}.

\bibliographystyle{aasjournal}
\bibliography{references_new}

\end{document}